\documentclass[11pt]{article}
\usepackage[margin=1in]{geometry}
\usepackage{fix-cm}
\usepackage{graphicx}
\usepackage{tikz,amssymb}
\usepackage{textcomp}
\usepackage{newtxtext}
\usepackage{newtxmath}
\usepackage[round]{natbib}
\usepackage{hyperref}
\usepackage{placeins}
\usepackage{xcolor}
\usepackage{amsmath,amsfonts,amssymb}
\usepackage{bm}
\usepackage{url}
\usepackage{color}
\usetikzlibrary{arrows.meta, decorations.pathreplacing, decorations.pathmorphing}

\usepackage{caption}
\newcommand{\yp}{{y^{+}}}
\newcommand{\dep}{{\delta^{+}}}
\newcommand{\LS}{\mathcal{L}}
\newcommand{\QS}{\mathcal{Q}}

\newcommand{\SSS}{\mathcal{S}}

\newcommand{\backsection}[2][]{\par\medskip\noindent\textbf{#1.}\enspace #2\par}

\title{\bfseries An Inner-Scaled Linear Contribution to Wall-Pressure
Variance at High Reynolds Number}

\author{%
  J. M. O. Massey\,$^{1}$\thanks{Email address for correspondence:
  \texttt{masseyj@stanford.edu}},\quad
  S. J. Zimmerman\,$^{2}$,\quad
  J. C. Klewicki\,$^{3}$,\quad
  B. J. McKeon\,$^{1}$ \\[6pt]
  {\small $^{1}$Center for Turbulence Research, Stanford University,
  Stanford, CA 94305, USA}\\
  {\small $^{2}$Department of Mechanical Engineering, Stony Brook University,
  Stony Brook, NY 11794, USA}\\
  {\small $^{3}$Department of Mechanical Engineering, University of Melbourne,
  Parkville, Victoria 3010, Australia}%
}
\date{}

\begin{document}
\maketitle

\begin{abstract}
    In canonical turbulent wall-bounded flows, the inner-scaled wall-pressure variance is empirically well described by a constant offset plus a slope logarithmic in the friction Reynolds number ($\dep$). Because the fluctuating pressure is predominantly a Poisson response to only two source terms---a linear contribution from the mean shear coupled to a fluctuating velocity gradient, and a nonlinear contribution from the fluctuating velocity field---the origin of this growth can be pinned down by elimination: if the linear source saturates at a Reynolds-number-independent value, the nonlinear source must carry the logarithmic growth. Here we supply the complementary evidence for inner-scaled invariance of the linear source at $\dep$ up to $O(10^4)$, using the simultaneous velocity and velocity-gradient hot-wire measurements of Zimmermann \textit{et al.} (2019 \textit{JFM}, vol. 869, pp. 182--213) acquired with a single eight-sensor probe in both a zero-pressure-gradient turbulent boundary layer and a high-Reynolds-number pipe flow. The inner-scaled factors entering the linear source collapse across Reynolds number, and the inertial-layer variance of the relevant fluctuating velocity gradient decays inversely with wall distance. Together with the established inner scaling of the mean shear, this is consistent with a linear wall-pressure contribution that, under inner normalisation, remains $O(1)$ as $\dep\to\infty$. Both source terms then trace to one structural mechanism: the near-wall depletion of mean spanwise vorticity that caps the linear source also feeds, through vortex stretching, the inertial-layer fissures that carry the growing nonlinear contribution.
\end{abstract}

\section{Introduction}\label{sec:intro}

    Canonical wall-bounded flows---pipes, channels, and zero-pressure-gradient boundary layers---share a layered inner-scaled architecture that is broadly universal across many of their statistical descriptors, with Reynolds-number dependence entering principally through the growing extent of the inertial interval or through outer corrections that vanish algebraically. In this context, wall-pressure fluctuations are a quantity whose inner-scaled variance does grow with Reynolds number: to good empirical approximation,
    \begin{equation}
        \langle p^2\rangle^+ = B + A\ln \dep,
        \label{eq:log_growth}
    \end{equation}
    where $p$ is the kinematic fluctuating wall pressure, $p^+ \equiv p/u_\tau^2$, $\langle p^2\rangle^+ \equiv \langle p^2\rangle/u_\tau^4$, $\langle\cdot\rangle$ denotes an ensemble or long-time average in a statistically stationary flow, and $\dep\equiv\delta u_\tau/\nu$ is the friction Reynolds number based on the characteristic outer length scale ($\delta=\delta_{99}$ for the boundary layer; the pipe radius $R$ for the pipe) \citep{farabee_spectral_1991,klewicki_statistical_2008,panton_correlation_2017,dacome_scaling_2025}. Given the universal inner-scaled structure of these flows, a natural question is whether the slope $A$ and the offset $B$ share a common mechanistic origin, or whether one source mechanism saturates at finite inner-scaled value and another supplies the $\ln\dep$ growth.

    Pressure is obtained from a Poisson equation and is therefore an elliptic response to source terms distributed across the flow; the wall signal is shaped jointly by the source structure and the Green's-function weighting that communicates it to the wall. With the velocity field split into mean and fluctuating parts, the fluctuating pressure--Poisson source admits a linear [$\LS$]/nonlinear [$\QS$] (a.k.a. rapid/slow) decomposition \citep{chou_velocity_1945,kim_structure_1989}, and within the nonlinear source the strain--vorticity form of \citet{bradshaw_note_1981} yields
    \begin{subequations}\label{eq:pp_vorticity}
    \begin{gather}
        \partial_i\partial_i \pi = \LS + \QS,\\
        \LS = 2 \Omega_z \partial_x v, \quad
        \QS = (\tfrac12 \omega_i\omega_i - s_{ij}s_{ij})-\overline{(\tfrac12 \omega_i\omega_i - s_{ij}s_{ij})},
    \end{gather}
    \end{subequations}
    where $\pi$ denotes the kinematic fluctuating pressure, $\Omega_z$ is the mean spanwise vorticity (in the pipe this is exactly $-{\rm d}U/{\rm d}y$, and approximately so in the TBL), $\omega_i$ is the fluctuating vorticity vector, and $s_{ij}$ is the fluctuating strain-rate tensor. The exact decomposition also contains a viscous (Stokes) contribution associated with no slip that becomes progressively less important with $\dep$ \citep{mansour_reynolds-stress_1988,kim_structure_1989,hoyas_reynolds_2008,gerolymos_wall_2013,panton_correlation_2017}. The wall-pressure variance therefore receives contributions from $\LS$, $\QS$ and the viscous source, together with their mutual cross-correlations.

    The linear source has been part of the wider wall-pressure discussion for decades. A textbook scaling argument predicts that the linear forcing should weaken with distance from the wall as the mean shear diminishes, while the strain--vorticity-form decomposition \eqref{eq:pp_vorticity} leads to interpretations that retain the centrality of mean-shear coupled contributions to the wall pressure \citep{bradshaw_turbulence_1967,bradshaw_note_1981}. The first fully resolved direct numerical simulation of channel flow at $\dep=180$ \citep{kim_turbulence_1987} and the associated pressure analysis \citep{kim_structure_1989} together established two especially relevant points. First, the nonlinear source $\QS$ is comparable to or larger than $\LS$ throughout the channel, contrary to the prior expectation that quadratic source terms should be negligible relative to the mean-shear-coupled linear term; the two become commensurate only very close to the wall, where $\LS$ remains a non-negligible contributor.
    Second, $\QS$ is itself the small residual of two large, strongly correlated quadratic forms ($\tfrac12\omega_i\omega_i$ and $s_{ij}s_{ij}$, equal on the average in homogeneous isotropic turbulence; \citealp{bradshaw_note_1981}), an observation reinforced by the source-component breakdown of \citet{kim_structure_1989} and revisited at the wall-mapping level by \citet{massey_behind_2025}.

    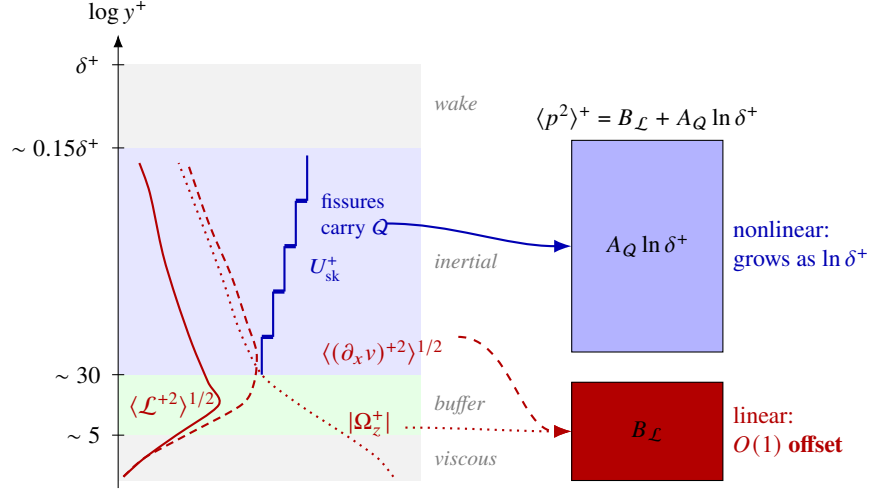
\begin{figure}
    \centering
    \begin{tikzpicture}[
    >=Latex, font=\fontsize{9pt}{11pt}\selectfont,
    layerlabel/.style={anchor=west, gray, font=\scriptsize\itshape}
]
\begin{scope}
\fill[gray!10] (0, 0)    rectangle (4, 0.6);  
\fill[green!10]  (0, 0.6)  rectangle (4, 1.4);  
\fill[blue!10] (0, 1.4)  rectangle (4, 4.4);  
\fill[gray!10] (0, 4.4)  rectangle (4, 5.5);  

\node[layerlabel]                at (4.05, 0.3) {viscous};
\node[layerlabel] (lbl-buffer)   at (4.05, 1.0) {buffer};
\node[layerlabel] (lbl-inertial) at (4.05, 2.9) {inertial};
\node[layerlabel]                at (4.05, 5.0) {wake};

\draw[->] (0, -0.1) -- (0, 5.9) node[above] {$\log y^+$};
\foreach \y/\lbl in {%
    0.6/{$\sim 5$}, 1.4/{$\sim 30$},
    4.4/{$\sim 0.15\delta^+$}, 5.5/{$\delta^+$}} {
    \draw (-0.08, \y) -- (0.08, \y);
    \node[left] at (-0.1, \y) {\lbl};
}

    \def\sx{1.9}
    \draw[blue!70!black, thick]
        (\sx,      1.40) -- (\sx,      1.90)
        (\sx+0.15, 1.90) -- (\sx+0.15, 2.50)
        (\sx+0.30, 2.50) -- (\sx+0.30, 3.10)
        (\sx+0.45, 3.10) -- (\sx+0.45, 3.70)
        (\sx+0.60, 3.70) -- (\sx+0.60, 4.30);
    \draw[blue!70!black, ultra thick]
        (\sx,      1.90) -- (\sx+0.15, 1.90)
        (\sx+0.15, 2.50) -- (\sx+0.30, 2.50)
        (\sx+0.30, 3.10) -- (\sx+0.45, 3.10)
        (\sx+0.45, 3.70) -- (\sx+0.60, 3.70);
    \node[blue!70!black, font=\scriptsize, anchor=west] at (\sx+0.5, 2.8) {$U^{+}_{\rm sk}$};
    \node[blue!70!black, font=\scriptsize, align=left, anchor=west] at (\sx+0.65, 3.5)
        {fissures\\carry $\QS$};

\draw[red!70!black, thick, dotted]
    plot[smooth] coordinates
    {(3.64, 0.05) (3.43, 0.30) (3.01, 0.60) (2.4, 1.00)
     (1.9, 1.40) (1.6, 2.00) (1.37, 2.70) (1.07, 3.50)
     (0.8, 4.20)};
\node[red!70!black, right] at (2.90, 0.75) {$|\Omega_z^+|$};

\draw[red!70!black, thick, densely dashed]
    plot[smooth] coordinates
    {(0.07, 0.05) (0.35, 0.30) (0.90, 0.60) (1.63, 1.00)
     (1.8, 1.30) (1.83, 1.70) (1.7, 2.10) (1.50, 2.70)
     (1.20, 3.40) (0.92, 4.20)};
\node[red!70!black, right] at (2.55, 1.70)
    {$\langle(\partial_x v)^{+2}\rangle^{1/2}$};

\draw[red!70!black, thick]
    plot[smooth] coordinates
    {(0.07, 0.05) (0.35, 0.30) (0.77, 0.60) (1.33, 1.00)
     (1.19, 1.30) (0.98, 1.80) (0.77, 2.40) (0.59, 3.00)
     (0.42, 3.80) (0.28, 4.20)};
\node[red!70!black, right] at (0, 1.00)
    {$\langle\mathcal{L}^{+2}\rangle^{1/2}$};

\end{scope}

\draw[->, thick, red!70!black, dotted] (3.8, 0.7) 
    to[out=0, in=180, looseness=0.9] (6, 0.65);
    
\draw[->, thick, red!70!black, dashed] (4.5, 1.9) 
    to[out=0, in=180, looseness=0.9] (6, 0.65);

\draw[->, thick, blue!70!black] (3.55, 3.4) 
    to[out=0, in=180, looseness=0.9] (6, 3.1);
\begin{scope}[xshift=6cm]
\fill[red!70!black]  (0, 0)   rectangle (2, 1.3);
\fill[blue!30] (0, 1.7) rectangle (2, 4.5);
\draw          (0, 0)   rectangle (2, 1.3);
\draw          (0, 1.7) rectangle (2, 4.5);

\node at (1, 0.65) {$B_{\mathcal{L}}$};
\node at (1, 3.10) {$A_{\mathcal{Q}}\ln\delta^+$};

\node[above] at (1, 4.5) {$\langle p^{2}\rangle^{+}$ = $B_{\mathcal{L}} + A_{\mathcal{Q}}\ln\delta^+$};
\node[red!70!black,  right, align=left] at (2, 0.65)
    {linear:\\ \textbf{$O(1)$ offset}};
\node[blue!70!black, right, align=left] at (2, 3.10)
    {nonlinear:\\ grows as $\ln\delta^+$};
\end{scope}
\end{tikzpicture}
    \caption{Schematic of the linear and nonlinear source contributions to wall pressure variance.
    \textit{Left:} under inner scaling, $|\Omega_z^+|$ drops from $1$ at the wall to $(\kappa y^+)^{-1}$ in the log region (red dashed); where $\kappa$ is the von K\'arm\'an constant. Under the working hypothesis tested here, $\langle(\partial_x v)^{+2}\rangle^{1/2}$ (red densely dashed) collapses across $\dep$, so the linear-source rms $\langle\LS^{+2}\rangle^{1/2}$ (red solid) peaks in the buffer and decays through the inertial layer. The blue stepped-profile sketches the skeletal staircase mean velocity $U^{\rm sk}$ of the UMZ--vortical-fissure model; the steps mark the fissures that act as compact carriers of $\QS$. \textit{Right:} the linear contribution to $\langle p^2\rangle^+$ saturates at the $\dep$-independent offset $B_\LS$, leaving the $\ln\dep$ growth (Eq.~\ref{eq:log_growth}) to be supplied by $\QS$.}
    \label{fig:schematic}
    \end{figure}

    Source-based analyses combine these source statistics with the elliptic Green's-function weighting, which attenuates contributions over wall-normal distances and wavenumber magnitudes \citep{anantharamu_analysis_2020,massey_behind_2025}. \citet{panton_correlation_2017} fitted both linear and nonlinear inner-scaled contributions to $\langle p^2\rangle^+$ in channel DNS as logarithmic in $\dep$ over $\dep\lesssim 5200$, with the nonlinear component the steeper, and thus increasingly more significant. \citet{massey_behind_2025} suggested that this logarithmic growth could be explained by the concentration of the nonlinear source term in inertial-layer vortical fissures, but the short inertial range available at that $\dep$ leaves the high-$\dep$ extrapolation an open empirical question. 
    
    This work (conceptually summarised in figure~\ref{fig:schematic}) refines that picture by supplying its linear side. Under the joint inner-scaling of $\Omega_z^+$ and $(\partial_x v)^+$, the linear-source r.m.s. is confined to the buffer layer, so that its integrated wall contribution saturates at $B_\LS$; the $\dep$ growth is left to $\QS$, carried by the inertial-layer vortical fissures that the same near-wall vorticity depletion seeds through vortex stretching \citep{meinhart_existence_1995,klewicki_description_2013,bautista_uniform_2019,massey_behind_2025}.

    The saturating contribution to $B_\LS$ rests on a claim less directly documented at low $\dep$: $(\partial_x v)^+$ must itself remain effectively inner-scaled, so that $\LS^+ = 2\Omega_z^+\,(\partial_x v)^+$ cannot acquire an additional $\dep$ dependence. Its wall-normal distribution and scale organisation cannot be inferred from wall-pressure or mean-velocity data, and DNS resolving the full pressure source remains confined to much lower $\dep$ than the highest laboratory pipe cases. We address this using the dataset of \citet{zimmerman_comparative_2019}: simultaneous measurements of all three velocity components and seven of the nine velocity-gradient components, acquired with a single eight-sensor velocity--vorticity hot-wire probe in both a zero-pressure-gradient turbulent boundary layer (Melbourne Wind Tunnel) and a high-Reynolds-number pipe (CICLoPE) up to $\dep\sim 10^4$. This matched-resolution pair across the two canonical geometries, with simultaneous access to the velocity-gradient components that enter $\LS$, provides exactly this high-$\dep$ access to the linear-source statistics. Anticipating the discussion (\S\,\ref{sec:structure}), the same near-wall vorticity budget ties the two logarithms together: the $\ln\dep$ in $\langle p^2\rangle^+$ and the $\ln\dep$ in the inner-scaled free-stream velocity $U_\infty^+$ count one and the same logarithmic depth of the inertial layer.

\section{Theoretical framework}\label{sec:framework}

    \subsection{Source-to-wall mapping}\label{sec:mapping}
    Solving the planar-Fourier-transformed pressure--Poisson problem for each source component yields the wall pressure as a Green's-function convolution over the source distribution. The key feature of this mapping, common to all canonical wall-bounded geometries, is that source contributions are exponentially attenuated over wall-normal distances $O(1/k)$, where $k=\sqrt{k_x^2+k_z^2}$ is the wall-parallel wavenumber magnitude \citep{anantharamu_analysis_2020,massey_behind_2025}. For the open boundary-layer half-space the wall kernel is
    \begin{equation}\label{eq:approx G decay}
        G_{\infty}(k,y') = -\frac{1}{k}{\rm e}^{-ky'},
    \end{equation}
    and the channel and pipe analogues replace the cross-stream operator with the appropriate finite-domain kernel but share the same exponential-attenuation structure. Wall-relevant source contributions therefore satisfy an efficient-communication envelope
    \begin{equation}
        \xi \equiv ky = O(1),
        \label{eq:envelope}
    \end{equation}
    in the sense that a source layer at wall-normal position $y$ couples most efficiently to wavelengths $k\sim 1/y$, and on the same envelope $|G|\sim 1/k\sim y$. Short wall-parallel wavelengths are therefore supplied predominantly from near-wall source layers, whereas long wavelengths sample farther into the flow; in inner scaling, outer-region sources project onto wall-parallel wavenumbers that vanish in the high-$\dep$ limit.
    This reduces the linear-source question to a diagnostic about the wall-relevant statistics of $\LS^+$, addressed in \S\,\ref{sec:working_hypothesis}.

    \subsection{Working hypothesis}\label{sec:working_hypothesis}
        The working hypothesis is that the linear wall-pressure contribution becomes $\dep$ invariant under inner scaling,
        \begin{equation}
            \langle p_{\LS}^{2} \rangle^+ \;\to\; B_{\LS},
            \qquad \text{as } \dep \to \infty,
        \label{eq:hyp_plateau}
        \end{equation}
        with algebraic finite-$\dep$ corrections. Equation~\eqref{eq:hyp_plateau} is a statement about the linear Poisson contribution itself, not the exact wall pressure.

        \citet{massey_behind_2025} verify \eqref{eq:hyp_plateau} at $\dep\approx 550$ by showing that the linear source density $\SSS_\LS^+$---so that $\langle p_\LS^2 \rangle^+ = \int_0^{\dep} \SSS_\LS^+(y^+)\,\mathrm{d}y^+$---decays through the inertial layer faster than the $\SSS_\LS^+\!\sim\!(y^+)^{-1}$ form required to generate a logarithmic contribution. Full-field source-density data are, however, not yet readily available at the high $\dep$ relevant to laboratory wall-pressure datasets. The complementary empirical question, accessible at high $\dep$, is whether the factors entering
        \begin{equation}
            \LS^+ \;=\; 2\Omega_z^+\,(\partial_x v)^+
        \label{eq:Lplus_def}
        \end{equation}
        remain inner-scaled, with $\Omega_z^+ \equiv \Omega_z\nu/u_\tau^2$ and $(\partial_x v)^+ \equiv (\nu/u_\tau^2)\,\partial_x v$.

        Following \citet{klewicki_description_2013}, $\Omega_z^+$ inner-scales to leading order, with $|\Omega_z^+| = 1$ at the wall and $|\Omega_z^+|\sim (\kappa y^+)^{-1}$ in the log region, where $\kappa$ is the von K\'arm\'an constant (figure~\ref{fig:schematic}, red dotted); interior to which half the inner-scaled circulation $U_\infty^+$ has already been laid down (cf.~\S\,\ref{sec:structure}).

        \begin{figure}
            \centering
            \includegraphics[width=0.9\linewidth]{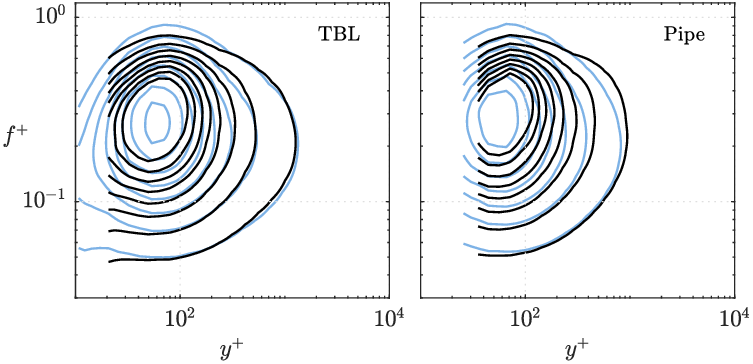}
            \caption{Contours of the premultiplied inner-scaled frequency spectrum $f^+\langle\phi_{21}^+\rangle$ of $(\partial_x v)^+$---where $\phi_{21}$ is the spectral density of the velocity-gradient component $\partial u_2/\partial x_1=\partial_x v$, so that $\int_0^\infty \langle\phi_{21}^+\rangle\,\mathrm{d}f^+=\langle(\partial_x v)^{+2}\rangle$---from the full premultiplied frequency spectra (supplementary figure~1) replotted on common axes for the lowest (light blue) and highest (black) $\dep$ cases: $\dep\approx 3300$ and $6300$ for the boundary layer (left), and $\dep\approx 5200$ and $10000$ for the pipe (right).}
        \label{fig:2-D atop}
        \end{figure}
        
        We begin with the zeroth-order diagnostic: the inner-scaled variance $\langle (\partial_x v)^{+2}\rangle$ (a sketch of the r.m.s. is shown in figure~\ref{fig:schematic}, red dashed). In the inertial layer, we parameterise it as
        \begin{equation}
            \langle (\partial_x v)^{+2}\rangle \sim (y^+)^{-\beta}.
        \end{equation}
        Which, together with $|\Omega_z^+|\sim (\kappa y^+)^{-1}$, gives
        \begin{equation}
            \langle (\LS^{+})^{2}\rangle \sim (y^+)^{-(2+\beta)}.
        \label{eq:beta_def}
        \end{equation}
        The relevance of this exponent to \eqref{eq:hyp_plateau} follows from a short envelope argument. On \eqref{eq:envelope}, with $|G^+|\sim y^+$ and one decade of efficiently contributing in-plane modes per decade of $y^+$, the per-decade source-density contribution to the linear wall-pressure variance scales as
        \begin{equation}
            y^+\SSS_\LS^+(y^+) \;\sim\; |G^+|^2\,\langle (\LS^{+})^{2}\rangle \;\sim\; (y^+)^{-\beta}.
        \label{eq:per_decade}
        \end{equation}
        Equation~\eqref{eq:per_decade} replaces the communicating-scale source spectral density that the kernel strictly samples with the total inner-scaled variance $\langle(\LS^+)^2\rangle$. Because $(\partial_x v)$ is a derivative, its variance is weighted towards scales shorter than the wall-coupling wavelength $k^+\sim 1/y^+$, so the two differ by a $y^+$-dependent spectral-shape factor. Two consequences follow with different robustness. The $\dep$-invariance of the wall contribution requires only that the inner-scaled spectral organisation of $(\partial_x v)^+$ be $\dep$-invariant, which figure~\ref{fig:2-D atop} supports directly. Identifying the per-decade exponent with $\beta$ additionally requires self-similarity of that organisation in $y^+$---the wall-blocking, attached-eddy picture---which the present spectra do not establish; we adopt it as a working assumption. The saturation conclusion is insensitive to the latter (any positive effective exponent suffices), whereas the predicted approach rate $(\dep)^{-\beta}$ is not. For any $\beta>0$, the integral $\int^{\dep}\!(y^+)^{-\beta}\,{\rm d}\ln y^+$ then converges as $\dep\to\infty$ and the wall contribution from $\LS$ approaches a finite asymptote $B_\LS$ with algebraic correction of order $(\dep)^{-\beta}$. The empirical task is therefore to verify, from the available high-$\dep$ data, that the factors entering \eqref{eq:Lplus_def} are inner-scaled and that the inertial-layer exponent $\beta$ lies above the marginal value $\beta=0$ required for saturation.

\section{Data collection and processing}\label{sec:data_processing}
        \begin{table}
        \centering
        \begin{tabular}{lcccccccc}
            Flow & $\delta^+$ & $\delta^+_{\mathrm{Chauhan}}$ & $u_\tau$ [m/s] &
            $\delta_\nu$ [\textmu m] & $l^+$ & $l_{wp}^+$ & Tier \\[2pt]
            \hline
            TBL  & 3300  & 4100 & 0.18 & 91 & 12 &  9 &  1 \\
            Pipe & 5200  & ---  & 0.18 & 87 & 13 &  9 &  1 \\
            TBL  & 4800  & 6000 & 0.25 & 63 & 18 & 12 &  2 \\
            Pipe & 7700  & ---  & 0.26 & 58 & 19 & 14 &  2 \\
            TBL  & 6300  & 8100 & 0.33 & 49 & 23 & 17 &  3 \\
            Pipe & 10000 & ---  & 0.34 & 45 & 25 & 18 &  3 \\
        \end{tabular}
        \caption{Experimental cases. $\delta^+$ uses $\delta_{99}$ for the boundary layer and the pipe radius $R$ for pipe; $\delta^+_{\mathrm{Chauhan}}$ is the boundary-layer thickness obtained from the inner--log--wake composite mean-velocity fit of \citet{chauhan_criteria_2009}. $l^+$ is the inner-normalised single-wire length; $l_{wp}^+$ is the wall-parallel probe footprint. Tiers indicate matched-resolution pipe/TBL pairings.}
        \label{tab:cases}
        \end{table}

        The data used here were originally reported in \cite{zimmerman_comparative_2019}. Probe and flow parameters are summarised in table~\ref{tab:cases}; the matched-resolution pairing across tiers is visible in the $l^+$ column. The cases comprise three ZPG boundary-layer measurements at $\delta^+\approx 3300$--$6300$ (equivalently, $\dep \approx 4100$--$8100$ using the composite-fit $\delta$ of \citealp{chauhan_criteria_2009}, which determines $\delta$ from the full inner--log--wake mean-velocity profile rather than from the local $U/U_\infty=0.99$ crossing and so removes the sensitivity of $\delta_{99}$ to facility-specific free-stream conditions) and three pipe measurements at $\delta^+\approx 5200$--$10000$. These data were acquired with the eight-sensor velocity--vorticity hot-wire probe described in \cite{zimmerman_design_2017} and post-processed according to \cite{zimmerman_experimental_2019}, in the Melbourne Wind Tunnel \citep{nickels_evidence_2005} and CICLoPE \citep{bellani_final_2016} respectively. The matched inner-normalised footprint within each tier means that any spatial-resolution effects (and the compensation applied for them) are approximately equivalent between the two flows (see $l^+$ and $l_{wp}^+$ columns in table~\ref{tab:cases}).
        
        This dataset resolves the velocity and velocity-gradient field from the buffer layer to the free stream/centreline. Across tiers the probe-to-$\delta_\nu$ ratio grows by $\sim70\%$, which is inconsequential for velocity statistics (single-wire lengths remain below $25\delta_\nu$ throughout) but must be compensated for in velocity-gradient spectra and variances. We adopt the synthetic-experiment compensation developed in \citet{zimmerman_design_2017} and applied in \citet{zimmerman_comparative_2019}: the velocity--vorticity probe is deployed in the boundary-layer DNS of \citet{sillero_one_2013} at the matching $l^+$, the effective amplitude filter $|G_{\rm eff}(k_1^+,y^+,l^+)|^2 \equiv E_{\rm meas}/E_{\rm act}$ (where $E_{\rm meas}$ and $E_{\rm act}$ are the measured and actual power spectral densities respectively) is determined for each component, and the experimental spectra are divided through by this filter at the corresponding wall distance. In the log and wake layers the recovered $|G_{\rm eff}|^2$ behaves close to the isotropic line-filter prediction of \citet{zhu_spatial_1995} (see also \citealp{pope_turbulent_2001}); the interpolation between the $\delta^+\!\approx\!2000$ DNS and the higher-$\delta^+$ experiments follows the wall-distance-dependent procedure detailed in \citet{zimmerman_experimental_2019}.

\section{Results}

\begin{figure}
        \centering
        \includegraphics[width=0.95\linewidth]{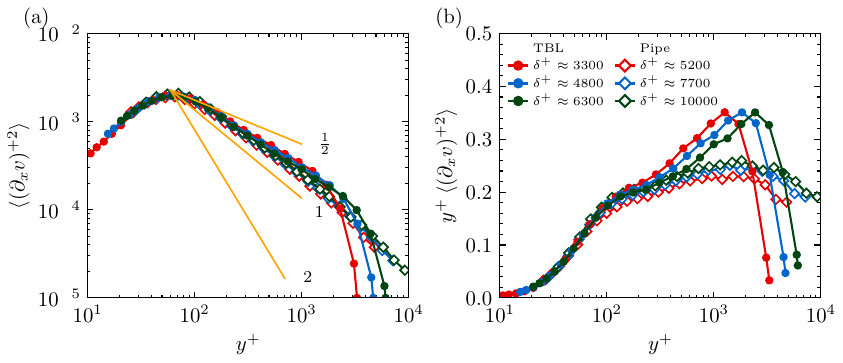}
        \caption{Inner-scaled variance profiles of $\langle (\partial_x v)^{+2}\rangle$ in the zero-pressure-gradient boundary layer and pipe flow. Panel (a) shows the log--log representation with a $-1/2$, $-1$, and $-2$ reference slope. Panel (b) shows the premultiplied form $y^+\langle (\partial_x v)^{+2}\rangle$.}
    \label{fig:BL_variance}
    \end{figure}

    Figure~\ref{fig:BL_variance}(a) shows that the profiles of $\langle (\partial_x v)^{+2}\rangle$ collapse well across Reynolds number through the near-wall region and the start of the inertial layer. Beyond the peak near $y^+\approx 60$, the decay is close to $(y^+)^{-1}$, with the higher-$\dep$ pipe cases providing the clearest extended range. Figure~\ref{fig:BL_variance}(b) is the more discriminating view. If the decay were exactly $(y^+)^{-1}$ over a given interval, the premultiplied curves would be flat there. The pipe data are close to that behaviour over a broad range, most clearly for the $\dep\approx 10^4$ case, and therefore provide the cleanest support for $\beta\simeq 1$. The ZPG boundary-layer curves are flatter only over a shorter interval and then rise gradually through the outer part of the inertial region before their geometric cutoff near $y^+=O(\dep)$.

    A further feature of figure~\ref{fig:BL_variance}(b) directly relevant to the inner-scaling claim is that the peak heights of $y^+\langle(\partial_x v)^{+2}\rangle$ are themselves approximately $\dep$ invariant in both flows, with only the abscissa location shifting: the pipe cases all reach a common peak magnitude on the inner-normalised ordinate, and the boundary-layer cases similarly collapse onto a common peak amplitude, with the peak migrating outward in $y^+$ as $\dep$ increases.
    This is the signature expected of an inner-scaled $\beta\simeq 1$ envelope: the local product $|\Omega_z^+|^2 \langle(\partial_x v)^{+2}\rangle$ that enters $\langle(\LS^+)^{2}\rangle$ does not gain amplitude with Reynolds number, so any $\dep$ dependence of the integrated wall contribution must come from changes in the extent of the inertial interval rather than from changes in the inner-scaled level itself.

    The upturn in the boundary-layer data, which begins once $\yp\gtrsim 0.07\dep$ (as verified by plotting on a $y/\delta$ abcissa), within the nominal inertial layer of figure~\ref{fig:schematic}, is most naturally interpreted as the onset of outer influence on the local source amplitude. The pipe data show only a slight incline, reflecting the weaker quiescent-core intermittency \citep{kwon_quiescent_2014}---the internal analogue of the TBL's turbulent/non-turbulent interface \citep[][\S5]{zimmerman_comparative_2019}---rather than its absence. The pipe's core interface borders weakly turbulent rather than irrotational fluid, so the vorticity contrast across it, and the associated intermittent large-scale gradient surplus, are correspondingly smaller. In both flows this surplus is a large-scale outer motion positioned at the far-from-wall, large-wavelength part of the source field, and is therefore suppressed both by the wall mapping \eqref{eq:envelope} and by the product with the decaying $\Omega_z^+\sim 1/\kappa y^+$; neither upturn threatens the linear-source argument. In the notation of \S\,\ref{sec:working_hypothesis}, the combined evidence therefore supports $\langle (\partial_x v)^{+2}\rangle\sim (y^+)^{-1}$ as the inertial-layer trend.

    Variance profiles alone do not fully determine the wall contribution, because the wall mapping in \eqref{eq:approx G decay} weights the source through its wall-parallel scale content. The available spectra do not provide the full $(k_x,k_z)$ dependence entering $k$, but they do test whether the dominant temporal and streamwise organisation of $(\partial_x v)^+$ changes materially with $\dep$. Figure~\ref{fig:2-D atop} provides evidence that they do not.

    The premultiplied frequency spectra retain essentially the same structure across Reynolds number in both geometries; the full set is provided as supplementary material. Figure~\ref{fig:2-D atop} overlays the lowest- and highest-$\dep$ contours on common axes. The dominant inner-scaled maximum remains centred near $y^+\approx 50$--$80$ and $f^+\approx 0.2$--$0.4$, with only a modest increase in peak magnitude at the highest $\dep$; on the plotted scale the change is less than $5\times 10^{-3}$, which is within the residual uncertainty of the resolution correction (\S\,\ref{sec:data_processing}). Integrated over $f^+$, this peak-magnitude difference accounts for a $\sim 15\%$ change in variance between the lowest- and highest-$\dep$ cases at $y^+\approx 50$, consistent with---though visually less striking than---the corresponding offset in the variance profiles of figure~\ref{fig:BL_variance}. The spectral peak is thus the footprint of the same weak residual $\dep$ trend rather than evidence of an outer, Reynolds-number-growing branch. If one interprets $f^+$ using a local Taylor hypothesis, this peak corresponds to convected streamwise wavelengths of only $30$--$70$ wall units at $y^+\approx 60$, consistent with $(\partial_x v)^+$ emphasising the shorter streamwise gradients associated with the near-wall cycle.

\section{Discussion}\label{sec:discussion}

\subsection{A saturating linear contribution}
    The principal result is not a new Reynolds-number fit but a mechanistic constraint, resting on the behaviour of $\langle (\partial_x v)^{+2}\rangle$ and the Green's-function weighting that maps it to the wall. The present experiments supply the missing linear-side evidence: not a direct decomposition of wall-pressure variance into linear and nonlinear parts, but high-$\dep$ access to the derivative statistics entering the linear source. The present measurements show an apparent Reynolds-number invariance of the inner-scaled variance $\langle (\partial_x v)^{+2}\rangle$ together with minimal structural change in the corresponding inner-scaled spectra. When combined with the established inner scaling of $\Omega_z^+$, this evidence is consistent with the linear source contributing a fixed $O(1)$ inner component to wall-pressure variance, rather than providing a substantial contribution to the term that grows with $\dep$ in $\langle p^2\rangle^+$ (cf. equation \ref{eq:log_growth}).

    The present conclusion is closely related to, but slightly stronger than, the empirical picture of \citet{panton_correlation_2017}, who fitted both the linear and nonlinear inner-scaled contributions to $\langle p^2 \rangle^+$ as separately logarithmic in $\delta^+$ over $\delta^+ \lesssim 5200$, with the nonlinear component the steeper. Over a bounded range, however, a saturating linear contribution and a weak logarithm are not interchangeable--the former concentrates its variation at low $\delta^+$ and flattens, whereas the latter rises by a fixed increment per decade--so we interpret Panton's fitted linear slope not as the true asymptote but as finite-$\delta^+$ phenomenology, a low $\dep$ feature that must ultimately turn over as $\langle p_\mathcal{L}^2 \rangle^+ \to B_\mathcal{L}$. The envelope argument introduced above fixes only the leading exponent of this approach, $(\delta^+)^{-\beta}$ with $\beta \simeq 1$, and not its finite-$\delta^+$ rate, which depends on the full linear source density; a logarithmic-looking transient over Panton's range is therefore not excluded. Because the present evidence constrains the factors entering $\mathcal{L}^+ = 2\Omega_z^+ (\partial_x v)^+$ rather than $\langle p_\mathcal{L}^2 \rangle^+$ itself, this saturation is a falsifiable prediction--a source-resolved computation of the linear wall-pressure variance at high-$\delta^+$ should reveal a departure from logarithmic growth--and the statement that $\mathcal{L}$ supplies only the $\delta^+$-independent offset is, in this sense, asymptotic: across Panton's range its fitted slope remains a non-negligible fraction of the total growth.

\subsection{The inertial-layer exponent $\beta\simeq 1$}
    The empirical exponent $\beta\simeq 1$ is, moreover, the value anticipated by classical inertial-layer scaling. In the locally isotropic dissipative range,
    \begin{equation}
        \langle (\partial_x v)^{+2}\rangle = \frac{2}{15}\,\varepsilon^+,
        \label{eq:isotropic_dxv}
    \end{equation}
    and the log region carries a dissipation profile $\varepsilon^+\sim 1/(\kappa y^+)$. The resulting prediction is
    \begin{equation}
        y^+\langle (\partial_x v)^{+2}\rangle \approx \frac{2}{15\kappa} \approx 0.34
        \quad \text{(for }\kappa\approx 0.39\text{)},
        \label{eq:kolmogorov_estimate}
    \end{equation}
    which sits within roughly a factor of $1.5$ of the compensated plateau values in figure~\ref{fig:BL_variance}(b)---reasonable given that local isotropy is approximate in the inertial layer of wall turbulence. The interest in the present finding is consequently not the value of $\beta$ itself, but that the data confirm it survives in the wall-relevant streamwise-derivative component across both canonical geometries, most cleanly in the highest-$\dep$ pipe case.

    The data also sharpen the rate at which the plateau in \eqref{eq:hyp_plateau} is approached. A naive envelope estimate---taking $v^+$ effectively constant on $\xi=ky=O(1)$, so that $(\partial_x v)^+\sim k^+v^+\sim (y^+)^{-1}$---would give $\langle(\partial_x v)^{+2}\rangle\sim (y^+)^{-2}$, hence $\beta=2$. The observed decay is slower, $\langle(\partial_x v)^{+2}\rangle\sim (y^+)^{-1}$ (so $\langle(\LS^+)^2\rangle\sim (y^+)^{-3}$, i.e.\ $\beta\simeq 1$), giving a finite-$\dep$ correction of order $(\dep)^{-1}$ rather than $(\dep)^{-2}$. This sets the rate at which the plateau is approached, not whether it is approached: the data remain consistent with a linear contribution that is asymptotically $O(1)$.

\subsection{A structural link between the linear and nonlinear sources}\label{sec:structure}
    A flow-physics reading runs alongside this elimination logic and links the two halves of the source decomposition through a single structural mechanism. Across the near-wall region (out to $y^+\simeq 50$), $|\Omega_z^+|$ drops precipitously from order unity at the wall toward its log-region $1/(\kappa y^+)$ form. Through the mean enstrophy equation, this depletion of mean spanwise vorticity is balanced by the production of fluctuating enstrophy via vortex stretching, driving the three-dimensionalisation and relative intensification of the fluctuating vorticity field. Farther from the wall, this reorganised vorticity concentrates into the inertial-layer vortical fissures sketched in figure~\ref{fig:schematic} \citep{klewicki_description_2013}, which act as compact carriers of $\QS$ \citep{massey_behind_2025}. The linear contribution isolated here and the nonlinear $\QS$ contribution are thus two faces of one structural story: the same near-wall $\Omega_z$ falloff that confines the integrated linear contribution to a $\dep$-independent inner envelope also energises the stretching that builds those fissures through which $\QS$ acquires its $\dep$-growing contribution.

    This structural link admits a compact quantitative reading. Inner normalisation fixes the wall as a source of mean spanwise vorticity of unit strength, $|\Omega_z^+|_{y^+=0}=1$---it is simply the inner-scaled wall shear---but it does not fix the total mean vorticity threaded through the layer. Integrating across the flow,
    \begin{equation}
        \underbrace{\int_0^{\dep}|\Omega_z^+|\,\mathrm{d}y^+}_{\Gamma_\infty} =  \underbrace{\int_{0}^{2.6\sqrt{\dep}}|\Omega_z^+|\,\mathrm{d}y^+}_{\sim\,\Gamma_\infty/2} + \underbrace{\int_{2.6\sqrt{\dep}}^{\dep}|\Omega_z^+|\,\mathrm{d}y^+}_{\sim\,\Gamma_\infty/2} = U_\infty^+ = \frac{1}{\kappa}\ln\dep + C,
        \label{eq:circulation}
    \end{equation}
    with $U_\infty^+$ the inner-scaled edge velocity---exactly the centreline velocity in the pipe, where $\Omega_z=-\mathrm{d}U/\mathrm{d}y$, and the free-stream velocity in the boundary layer. The inner-scaled circulation therefore grows logarithmically: the fixed-strength wall source feeds a $1/(\kappa y^+)$ mean-vorticity tail stretched over an inertial interval whose inner-scaled depth grows with $\dep$. Where this circulation is laid down is equally telling: since the inertial layer sets in at $\yp\simeq 2.6\sqrt{\dep}$ \citep{klewicki_description_2013} and $\ln\sqrt{\dep}=\tfrac12\ln\dep$, the circulation accumulated below the onset, $U^+(2.6\sqrt{\dep})$, carries exactly half the logarithmic slope of $U_\infty^+$, so that $U^+(2.6\sqrt{\dep})/U_\infty^+\to\tfrac12$ as $\dep\to\infty$---half the circulation is threaded below the onset and half above. The near-wall region that carries the linear source is correspondingly marginalised: the buffer occupies a vanishing fraction $\sim\dep^{-1/2}$ of the distance to the onset, and the circulation out to any fixed $\yp$ is an $O(1)$ increment that becomes a diminishing share of the logarithmically growing total. The linear source thus resides precisely where the circulation fraction vanishes, while the growth is spread over the inertial layer that houses the fissures. The mean vorticity depleted across this lengthening range feeds the fluctuating enstrophy that populates the fissures. Crucially, this conversion proceeds at a fixed inner-scaled local intensity---the $\dep$-invariant peak of $y^+\langle(\partial_x v)^{+2}\rangle$ in figure~\ref{fig:BL_variance}(b) is the linear-side evidence of that invariance---so the nonlinear contribution accumulates by extent rather than by intensification: more fissures spanning a longer $\yp$ staircase, not stronger ones. What permits a fixed-strength wall source to feed a growing wall-pressure contribution is therefore not a more efficient concentration of vorticity at higher $\dep$, but the logarithmic lengthening of the inertial interval over which an inner-invariant source is distributed; the $\ln\dep$ in $\langle p^2\rangle^+$ and the $\ln\dep$ in $U_\infty^+$ are the same count of the inertial layer's logarithmic depth.

\subsection{Consequences of the linear-source saturation}
    The consequence is therefore sharper than a generic constraint. The linear source is not negligible: the present evidence supports interpreting it as part of the fixed inner contribution to $\langle p^2\rangle^+$. Given the observed variance scaling of $(\partial_x v)^+$ and the known behaviour of $\Omega_z^+$, it is rationally implausible to reconcile the observed logarithmic growth of $\langle p^2\rangle^+$ with the linear term alone. The more plausible interpretation is that the growing contribution is supplied primarily by the nonlinear source $\QS$ and by the inertial-layer organisation that allows it to accumulate with Reynolds number. In that sense, the present result does more than remain compatible with the broader source-based picture: it strengthens the case for it.

\backsection[Supplementary material]{Supplementary material is available online and presents the full set of inner-scaled premultiplied frequency spectra of $(\partial_x v)^+$ for the boundary-layer and pipe cases, from which the common-axis overlay of figure~\ref{fig:2-D atop} is constructed.}
\backsection[Funding]{The support from DARPA under award HR0011-24-9-0465 and the support of ONR to CTR under grant N000142312833 is gratefully acknowledged.}
\backsection[Declaration of interests]{The authors report no conflict of interest.}

\bibliographystyle{jfm}
\bibliography{references}

@article{klewicki_statistical_2008,
	title = {Statistical structure of the fluctuating wall pressure and its in-plane gradients at high {Reynolds} number},
	volume = {609},
	copyright = {https://www.cambridge.org/core/terms},
	issn = {0022-1120, 1469-7645},
	url = {https://www.cambridge.org/core/product/identifier/S0022112008002541/type/journal_article},
	doi = {10.1017/S0022112008002541},
	language = {en},
	urldate = {2024-09-06},
	journal = {J. Fluid Mech.},
	author = {Klewicki, J. C. and Priyadarshana, P. J. A. and Metzger, M. M.},
	month = aug,
	year = {2008},
	pages = {195--220},
}

@article{farabee_spectral_1991,
	title = {Spectral features of wall pressure fluctuations beneath turbulent boundary layers},
	volume = {3},
	issn = {0899-8213},
	url = {https://pubs.aip.org/pof/article/3/10/2410/402018/Spectral-features-of-wall-pressure-fluctuations},
	doi = {10.1063/1.858179},
	language = {en},
	urldate = {2024-09-09},
	journal = {Phys. Fluids A},
	author = {Farabee, Theodore M. and Casarella, Mario J.},
	month = oct,
	year = {1991},
	pages = {2410--2420},
}

@article{kim_structure_1989,
	title = {On the structure of pressure fluctuations in simulated turbulent channel flow},
	volume = {205},
	issn = {0022-1120, 1469-7645},
	url = {http://www.journals.cambridge.org/abstract_S0022112089002090},
	doi = {10.1017/S0022112089002090},
	language = {en},
	urldate = {2024-12-05},
	journal = {J. Fluid Mech.},
	author = {Kim, John},
	month = aug,
	year = {1989},
	pages = {421},
}

@article{panton_correlation_2017,
	title = {Correlation of pressure fluctuations in turbulent wall layers},
	volume = {2},
	copyright = {https://link.aps.org/licenses/aps-default-license},
	issn = {2469-990X},
	url = {https://link.aps.org/doi/10.1103/PhysRevFluids.2.094604},
	doi = {10.1103/PhysRevFluids.2.094604},
	language = {en},
	urldate = {2025-02-04},
	journal = {Phys. Rev. Fluids},
	author = {Panton, Ronald L. and Lee, Myoungkyu and Moser, Robert D.},
	month = sep,
	year = {2017},
	pages = {094604},
}

@article{anantharamu_analysis_2020,
	title = {Analysis of wall-pressure fluctuation sources from direct numerical simulation of turbulent channel flow},
	volume = {898},
	copyright = {https://www.cambridge.org/core/terms},
	issn = {0022-1120, 1469-7645},
	url = {https://www.cambridge.org/core/product/identifier/S0022112020004127/type/journal_article},
	doi = {10.1017/jfm.2020.412},
	language = {en},
	urldate = {2025-02-20},
	journal = {J. Fluid Mech.},
	author = {Anantharamu, Sreevatsa and Mahesh, Krishnan},
	month = sep,
	year = {2020},
	pages = {A17},
}

@article{mansour_reynolds-stress_1988,
	title = {Reynolds-stress and dissipation-rate budgets in a turbulent channel flow},
	volume = {194},
	issn = {0022-1120, 1469-7645},
	url = {http://www.journals.cambridge.org/abstract_S0022112088002885},
	doi = {10.1017/S0022112088002885},
	language = {en},
	urldate = {2025-03-17},
	journal = {J. Fluid Mech.},
	author = {Mansour, N. N. and Kim, J. and Moin, P.},
	month = sep,
	year = {1988},
	pages = {15--44},
}

@article{klewicki_description_2013,
	title = {A description of turbulent wall-flow vorticity consistent with mean dynamics},
	volume = {737},
	copyright = {https://www.cambridge.org/core/terms},
	issn = {0022-1120, 1469-7645},
	url = {https://www.cambridge.org/core/product/identifier/S002211201300565X/type/journal_article},
	doi = {10.1017/jfm.2013.565},
	language = {en},
	urldate = {2025-06-13},
	journal = {J. Fluid Mech.},
	author = {Klewicki, J. C.},
	month = dec,
	year = {2013},
	pages = {176--204},
}

@article{bautista_uniform_2019,
	title = {A uniform momentum zone-vortical fissure model of the turbulent boundary layer},
	volume = {858},
	copyright = {https://www.cambridge.org/core/terms},
	issn = {0022-1120, 1469-7645},
	url = {https://www.cambridge.org/core/product/identifier/S0022112018007693/type/journal_article},
	doi = {10.1017/jfm.2018.769},
	language = {en},
	urldate = {2025-07-21},
	journal = {J. Fluid Mech.},
	author = {Bautista, Juan Carlos Cuevas and Ebadi, Alireza and White, Christopher M. and Chini, Gregory P. and Klewicki, Joseph C.},
	month = jan,
	year = {2019},
	pages = {609--633},
}

@article{dacome_scaling_2025,
	title = {Scaling of wall-pressure-velocity correlations in high-{Reynolds}-number turbulent pipe flow},
	volume = {1013},
	issn = {0022-1120, 1469-7645},
	url = {https://www.cambridge.org/core/product/identifier/S0022112025004008/type/journal_article},
	doi = {10.1017/jfm.2025.400},
	language = {en},
	urldate = {2025-07-29},
	journal = {J. Fluid Mech.},
	author = {Dacome, Giulio and Lazzarini, Lorenzo and Talamelli, Alessandro and Bellani, Gabriele and Baars, Woutijn J.},
	month = jun,
	year = {2025},
	pages = {A48},
}

@article{gerolymos_wall_2013,
	title = {Wall effects on pressure fluctuations in turbulent channel flow},
	volume = {720},
	copyright = {https://www.cambridge.org/core/terms},
	issn = {0022-1120, 1469-7645},
	url = {https://www.cambridge.org/core/product/identifier/S0022112012006337/type/journal_article},
	doi = {10.1017/jfm.2012.633},
	language = {en},
	urldate = {2025-08-29},
	journal = {J. Fluid Mech.},
	author = {Gerolymos, G. A. and Sénéchal, D. and Vallet, I.},
	month = apr,
	year = {2013},
	pages = {15--65},
}

@article{meinhart_existence_1995,
	title = {On the existence of uniform momentum zones in a turbulent boundary layer},
	volume = {7},
	issn = {1070-6631, 1089-7666},
	url = {https://pubs.aip.org/pof/article/7/4/694/258935/On-the-existence-of-uniform-momentum-zones-in-a},
	doi = {10.1063/1.868594},
	language = {en},
	urldate = {2025-10-09},
	journal = {Phys. Fluids},
	author = {Meinhart, Carl D. and Adrian, Ronald J.},
	month = apr,
	year = {1995},
	pages = {694--696},
}

@article{bradshaw_note_1981,
	title = {A note on {Poisson}’s equation for pressure in a turbulent flow},
	volume = {24},
	issn = {0031-9171},
	url = {https://pubs.aip.org/pfl/article/24/4/777/824022/A-note-on-Poisson-s-equation-for-pressure-in-a},
	doi = {10.1063/1.863442},
	language = {en},
	number = {4},
	urldate = {2025-12-10},
	journal = {The Physics of Fluids},
	author = {Bradshaw, P. and Koh, Y. M.},
	month = apr,
	year = {1981},
	pages = {777--777},
}

@phdthesis{zimmerman_experimental_2019,
	title = {Experimental investigation of velocity and vorticity in turbulent wall flows},
	language = {en},
	school = {University of Melbourne},
	author = {Zimmerman, Spencer James},
	year = {2019},
}

@misc{massey_behind_2025,
	title = {On the Poisson-Source Basis of Logarithmic Wall-Pressure-Variance Growth},
	url = {http://arxiv.org/abs/2511.16776},
	doi = {10.48550/arXiv.2511.16776},
	urldate = {2026-01-14},
	publisher = {arXiv},
	author = {Massey, Jonathan M. O. and Klewicki, Joseph C. and McKeon, Beverley J.},
	month = dec,
	year = {2026},
	note = {arXiv:2511.16776 [physics]},
}

@article{kwon_quiescent_2014,
	title = {The quiescent core of turbulent channel flow},
	volume = {751},
	doi = {10.1017/jfm.2014.295},
	journal = {J. Fluid Mech.},
	author = {Kwon, Y. S. and Philip, J. and de Silva, C. M. and Hutchins, N. and Monty, J. P.},
	year = {2014},
	pages = {228--254},
}

@article{kim_turbulence_1987,
	title = {Turbulence statistics in fully developed channel flow at low {Reynolds} number},
	volume = {177},
	copyright = {https://www.cambridge.org/core/terms},
	issn = {0022-1120, 1469-7645},
	url = {https://www.cambridge.org/core/product/identifier/S0022112087000892/type/journal_article},
	doi = {10.1017/S0022112087000892},
	language = {en},
	urldate = {2026-02-16},
	journal = {J. Fluid Mech.},
	author = {Kim, John and Moin, Parviz and Moser, Robert},
	month = apr,
	year = {1987},
	pages = {133--166},
}

@article{bradshaw_turbulence_1967,
	title = {The turbulence structure of equilibrium boundary layers},
	volume = {29},
	copyright = {https://www.cambridge.org/core/terms},
	issn = {0022-1120, 1469-7645},
	url = {https://www.cambridge.org/core/product/identifier/S0022112067001089/type/journal_article},
	doi = {10.1017/S0022112067001089},
	language = {en},
	number = {4},
	urldate = {2026-02-16},
	journal = {J. Fluid Mech.},
	author = {Bradshaw, P.},
	month = sep,
	year = {1967},
	pages = {625--645},
}

@article{zimmerman_comparative_2019,
	title = {A comparative study of the velocity and vorticity structure in pipes and boundary layers at friction {Reynolds} numbers up to},
	volume = {869},
	copyright = {https://www.cambridge.org/core/terms},
	issn = {0022-1120, 1469-7645},
	url = {https://www.cambridge.org/core/product/identifier/S0022112019001824/type/journal_article},
	doi = {10.1017/jfm.2019.182},
	language = {en},
	urldate = {2026-02-17},
	journal = {J. Fluid Mech.},
	author = {Zimmerman, S. and Philip, J. and Monty, J. and Talamelli, A. and Marusic, I. and Ganapathisubramani, B. and Hearst, R. J. and Bellani, G. and Baidya, R. and Samie, M. and Zheng, X. and Dogan, E. and Mascotelli, L. and Klewicki, J.},
	month = jun,
	year = {2019},
	pages = {182--213},
}

@article{chauhan_criteria_2009,
  title={Criteria for assessing experiments in zero pressure gradient boundary layers},
  author={Chauhan, K. A. and Monkewitz, P. A. and Nagib, H. M.},
  journal={Fluid Dyn. Res.},
  volume={41},
  number={2},
  pages={021404},
  year={2009}
}

@article{zimmerman_design_2017,
  title={Design and implementation of a hot-wire probe for simultaneous velocity and vorticity vector measurements in boundary layers},
  author={Zimmerman, S. and Morrill-Winter, C. and Klewicki, J.},
  journal={Exp. Fluids},
  volume={58},
  number={10},
  pages={148},
  year={2017},
  publisher={Springer}
}

@article{nickels_evidence_2005,
  title={Evidence of the $k_1^{-1}$ law in a high-{R}eynolds-number turbulent boundary layer},
  author={Nickels, T. B. and Marusic, I. and Hafez, S. and Chong, M. S.},
  journal={Phys. Rev. Lett.},
  volume={95},
  number={7},
  pages={074501},
  year={2005},
  publisher={APS}
}

@inproceedings{bellani_final_2016,
  title={The final design of the long pipe in {CICLOPE}},
  author={Bellani, G. and Talamelli, A.},
  booktitle={Prog. Turbul. VI, Proc. iTi Conf. Turbul. 2014},
  pages={205--209},
  year={2016},
  organization={Springer}
}

@article{sillero_one_2013,
  title={One-point statistics for turbulent wall-bounded flows at {R}eynolds numbers up to $\delta^+\approx 2000$},
  author={Sillero, J. A. and Jim{\'e}nez, J. and Moser, R. D.},
  journal={Phys. Fluids},
  volume={25},
  number={10},
  year={2013},
  publisher={AIP Publishing}
}

@book{pope_turbulent_2001,
  title={Turbulent Flows},
  publisher={Cambridge Univ. Press},
  author={Pope, S. B.},
  year={2000}
}

@article{zhu_spatial_1995,
  title={The spatial resolution of two {X}-probes for velocity derivative measurements},
  author={Zhu, Y. and Antonia, R. A.},
  journal={Meas. Sci. Tech.},
  volume={6},
  number={5},
  pages={538--549},
  year={1995}
}

@article{chou_velocity_1945,
  title   = {On velocity correlations and the solutions of the equations of turbulent fluctuation},
  author  = {Chou, P. Y.},
  journal = {Q. Appl. Math.},
  volume  = {3},
  number  = {1},
  pages   = {38--54},
  year    = {1945}
}

@article{hoyas_reynolds_2008,
  title   = {{R}eynolds number effects on the {R}eynolds-stress budgets in turbulent channels},
  author  = {Hoyas, S. and Jim{\'e}nez, J.},
  journal = {Phys. Fluids},
  volume  = {20},
  number  = {10},
  pages   = {101511},
  year    = {2008}
}

\end{document}